\documentclass{Interspeech2024}
\usepackage{hyperref}
\usepackage{makecell}
\usepackage{amssymb}
\usepackage{pifont}

\interspeechcameraready

\title{VoxSim: A perceptual voice similarity dataset}

\name[affiliation={1}]{Junseok}{Ahn}
\name[affiliation={1}]{Youkyum}{Kim}
\name[affiliation={2}]{Yeunju}{Choi}
\name[affiliation={1}]{Doyeop}{Kwak}
\name[affiliation={1}]{Ji-Hoon}{Kim}
\name[affiliation={2}]{Seongkyu}{Mun}
\name[affiliation={1}]{Joon~Son}{Chung}

\address{
  $^1$Korea Advanced Institute of Science and Technology, South Korea\\
  $^2$Samsung Research, South Korea}
\email{junseok.ahn@kaist.ac.kr}


\begin{document}

\maketitle

\begin{abstract}
    
    This paper introduces VoxSim, a dataset of perceptual voice similarity ratings. Recent efforts to automate the assessment of speech synthesis technologies have primarily focused on predicting mean opinion score of naturalness, leaving speaker voice similarity relatively unexplored due to a lack of extensive training data. To address this, we generate about 41k utterance pairs from the VoxCeleb dataset, a widely utilised speech dataset for speaker recognition, and collect nearly 70k speaker similarity scores through a listening test. VoxSim offers a valuable resource for the development and benchmarking of speaker similarity prediction models. We provide baseline results of speaker similarity prediction models on the VoxSim test set and further demonstrate that the model trained on our dataset generalises to the out-of-domain VCC2018 dataset. 

\end{abstract}

\section{Introduction}
\blfootnote{The dataset is available from}
\blfootnote{\url{https://mm.kaist.ac.kr/projects/voxsim}}

In many areas of machine learning, the objective is to train models that emulate human cognitive abilities and aim to match human-level performance~\cite{jordan2015machine,lecun2015deep}. However, there are cases where AI has outperformed human capabilities~\cite{chung2017lip,kim2023let}. Speaker recognition is a notable domain where AI models have shown superiority over human abilities. ECAPA-TDNN~\cite{desplanques2020ecapa} represents significant progress in the field of speaker verification, demonstrating an Equal Error Rate (EER) of less than 1\% on the VoxCeleb benchmark dataset~\cite{nagrani17voxceleb}. Recently, self-supervised learning-based models have shown even superior verification performance~\cite{chen2022large, chen2022wavlm}. In contrast, human performance in speaker identification falls considerably short. Huh et al.~\cite{kang2022augmentation} report that Amazon Mechanical Turk crowdworkers achieve an EER of 26.51\% on the same dataset, and even expert researchers in speaker recognition demonstrate an EER of 15.77\%. This indicates a substantial gap between the speaker characteristics that speaker recognition systems can extract and what humans can discern.

This gap leads to several issues, particularly when evaluating the speaker similarity of synthesised speech using speaker recognition systems. 
One common goal of speech generative models is to produce a consistent voice that closely matches a reference voice~\cite{casanova2022yourtts,shen2023naturalspeech,choi2022nansy++,choi2023diffv2s}.
Therefore, part of the evaluation of these systems is to verify the speaker similarity between the synthesised speech and the reference voice. An objective verification method involves extracting speaker feature embeddings from both voices using a speaker recognition model and measuring their cosine similarity~\cite{kim2023crossspeech, le2024voicebox, kim2024clam,jang2024faces}. However, this score often significantly deviates from what humans perceive~\cite{hu2022svsnet,deja22automatic}. As a result, the evaluation of synthesised speech relies heavily on subjective evaluation, a process that requires considerable time and resources.

\begin{table}[!t]
    \centering
    \caption{Data statistics for VCC2018, internal dataset from Deja et al.~\cite{deja22automatic}, and our VoxSim. {\bf \# spks.}: Total number of speakers. {\bf \# pairs}: Total number of utterance pairs. {\bf ratings}: Total number of similarity ratings. {\bf Unseen test spks.}: Whether the speakers in the test split were unseen during training. {\bf Public}: Whether the dataset is public.}
    \vspace{-2mm}
    \resizebox{\linewidth}{!}{
        \begin{tabular}{cccccc}
        \toprule
        \textbf{Dataset} & \textbf{\# spks.} & \textbf{\# pairs} & \textbf{\# ratings} & \makecell{\textbf{Unseen} \\ \textbf{test spks.}} & \textbf{Public} \\
        \midrule
        VCC2018 & 12 & 21,562 & 30,864 & \ding{55} & \ding{51} \\
        Deja et al.~\cite{deja22automatic} & 13 & 18,493 & 730,308 & \ding{55} & \ding{55} \\
        \midrule
        VoxSim & 1,251 & 41,578 & 69,409 & \ding{51} & \ding{51} \\
        \bottomrule
        \end{tabular}
    }
    \vspace{-2mm}
    \label{tab:data_compare}
\end{table}
To address this problem, techniques to automate speaker similarity assessment for synthetic speech are needed but have not been well-explored. To the best of our knowledge, there have been only two attempts at this automation. SVSNet~\cite{hu2022svsnet} is the first end-to-end neural network model designed to evaluate speaker similarity between converted and natural speech in a voice conversion task. SVSNet takes raw waveforms as input rather than using handcrafted features to analyse speech more accurately and introduces a co-attention mechanism to resolve length and content mismatches between the two voices. This structure achieves an utterance-level linear correlation coefficient of 0.574 on the VCC2018~\cite{lorenzotrueba18_odyssey} speaker similarity evaluation dataset, released by the voice conversion challenge. Deja et al.~\cite{deja22automatic} propose an automated method to evaluate speaker similarity by extending the speaker verification system. They synthesise speech samples using 354 modern text-to-speech systems and collect MUSHRA scores~\cite{series2014method} from listening tests to build their own dataset. The authors build a regression model to predict MUSHRA speaker similarity scores from speaker embeddings of two utterances and propose a loss to compensate for data imbalance. 

\begin{figure*}[!t]
   \centering
   \setlength{\tabcolsep}{7pt}
   \begin{tabular}{cccc}
        \includegraphics[height=2.6cm]{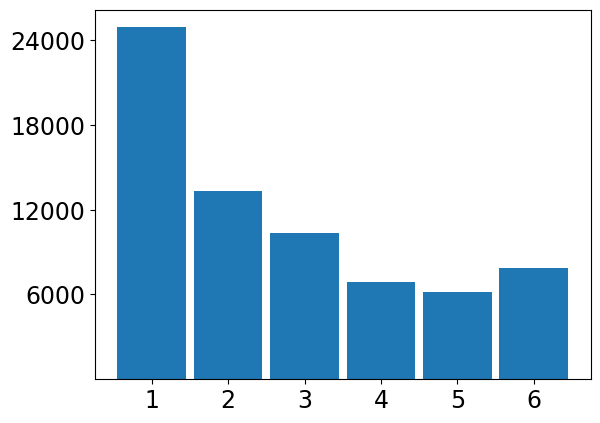} &
        \includegraphics[height=2.6cm]{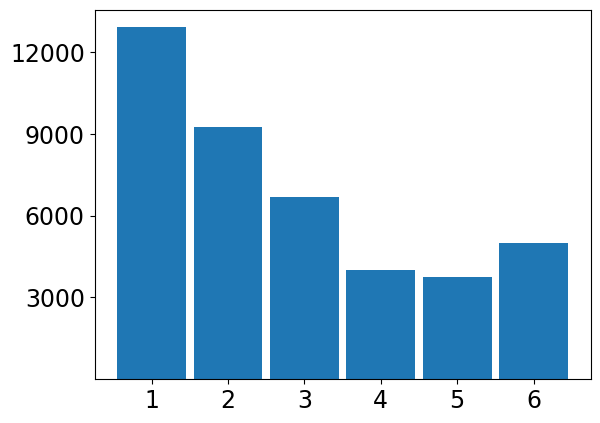} &
        \includegraphics[height=2.6cm]{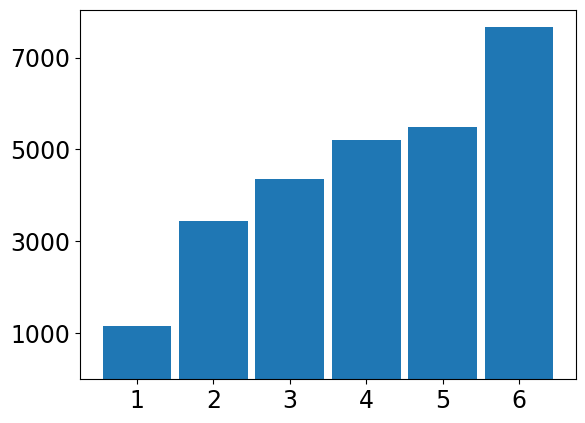} &
        \includegraphics[height=2.6cm]{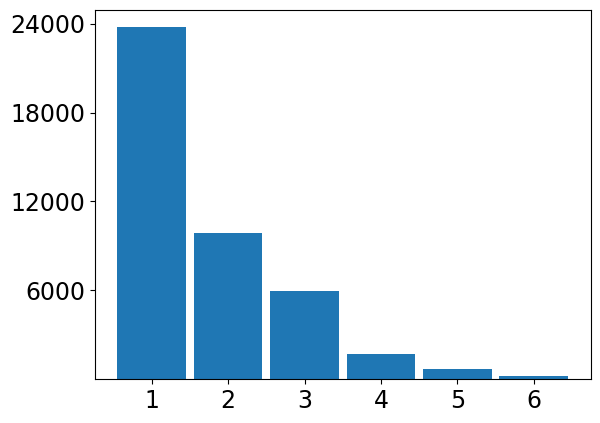} \\
        (a) individual scores & (b) average scores & (c) same speaker pairs & (d) different speaker pairs  \\
   \end{tabular}
   \vspace{-2mm}
   \caption{Distributions of (a) individual scores, (b) average scores per utterance pair, (c) scores for the same speaker pairs, and (d) scores for different speaker pairs. The x-axis represents the score label and the y-axis represents the number of ratings.}
   \vspace{-4mm}
   \label{fig:sdist}
\end{figure*}

The primary challenge in developing automated speech evaluation models is the scarcity of public data~\cite{reddy2021dnsmos,maiti2023speechlmscore}. The VCC2018 dataset, utilised for training SVSNet, comprises 30,864 speaker similarity scores for 21,562 pairs of converted and natural utterances from 36 voice conversion systems. Each pair is evaluated by 1 to 8 subjects through crowdsourced listening tests. In Deja et al.'s study, the model is trained and evaluated on data collected by the authors, which has not been made publicly available. The training data are limited not only in size but also in the diversity of speakers, with both datasets featuring a very small number of speakers. The VCC2018 dataset includes 12 speakers and 32 speaker combinations, whereas the other dataset contains only 13 target speakers. This limitation suggests that the trained similarity prediction models may not perform well on speech pairs from unseen speakers and are not suitable for evaluating zero-shot systems~\cite{jia2018transfer,arik2018neural}. Moreover, both datasets only consist of pairs of natural speech and speech synthesised by a few selected speech synthesis systems, indicating a lack of generalisability across different domains.

In this work, we introduce VoxSim, a large-scale open dataset that evaluates cognitive speaker similarity scores for speech pairs. To the best of our knowledge, this is the first dataset specifically crafted for training models that automate voice similarity assessment. VoxSim consists of approximately 70k similarity ratings from over 1k speakers. Since the utterances are sampled from VoxCeleb1~\cite{nagrani17voxceleb}, the evaluation model is exposed to a variety of channel effects and noise during training, enhancing its generalisation performance across speech domains. Table~\ref{tab:data_compare} compares VoxSim with the previous two datasets in terms of the number of speakers, pairs, ratings, the unseen status of test speakers, and their availability to the public. We provide baseline results for various speaker similarity prediction models on the VoxSim dataset and demonstrate the generalisability of our data through testing on the VCC2018 dataset.
\section{VoxSim Dataset}

\begin{figure}[!b]
   \centering
   \vspace{-3mm}
   \includegraphics[width=1.0\linewidth]{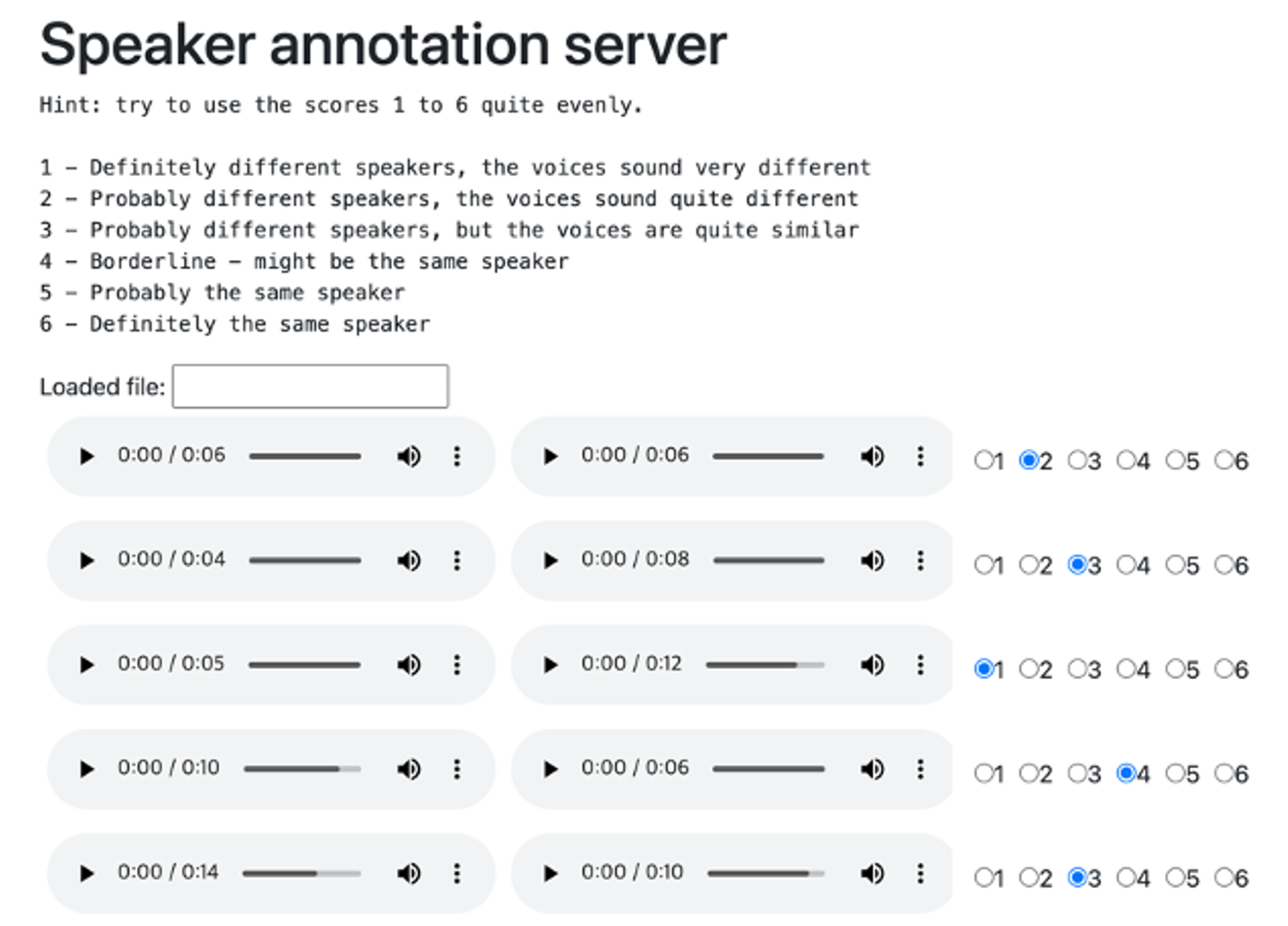}
   \vspace{-3mm}
   \caption{Speaker similarity annotation page.}
   \label{fig:annotation}
\end{figure}

\vspace{3pt}\noindent\textbf{Data source.} 
Speaker similarity ratings are collected using utterances from VoxCeleb1, which serves as the benchmark dataset for speaker identification and verification tasks. It consists of utterances extracted from videos uploaded to YouTube, thus the speech segments contain various acoustic environments. The speakers encompass a wide range of nationalities and ages. We create 50k random utterance pairs and conduct a listening test to evaluate the voice similarity. Diverse speaker combinations are created by organising speaker pairs to ensure a 1.5 times higher number of different speaker pairs compared to the same speaker pairs.

\vspace{3pt}\noindent\textbf{Listening test procedure.} 
Twelve evaluators participate in this listening test and they are asked to evaluate the speaker similarity of each pair on a 6-point Likert scale; {\it 1: Definitely different speakers, 2: Probably different speakers, 3: Possibly different speakers, 4: Possibly the same speaker, 5: Probably the same speaker, and 6: Definitely the same speaker}. Evaluators are requested to rate as evenly as possible from 1 to 6 points, aiming to prevent the dataset's score distribution from being skewed towards either 1 or 6. 
Given the diverse environmental and linguistic contexts present in VoxCeleb dataset, the evaluators are guided to focus solely on the voice characteristics of the main speaker, regardless of the content of the utterance, language, and the acoustic environment. An example annotation page is shown in Fig.~\ref{fig:annotation}.

\vspace{3pt}\noindent\textbf{Quality control.} 
During the listening test, the quality of the collected scores is controlled through periodic reviews of the speaker verification rate. Min-max normalisation is utilised to project the scores between 0 and 1, enabling the calculation of the EER against the actual speaker labels. If an evaluator's EER significantly deviates from the average EER of all evaluators, we provide feedback and ask the evaluator to conduct a re-assessment.
At the end of the collection, the average speaker verification rate for all evaluators is an EER of 17.7\%. After the listening test, pairs with a difference of more than 3 points between the highest and lowest evaluation scores are considered outliers and excluded.

\begin{table}[!t]
    \centering
    \caption{VoxSim Train and Test set statistics. {\bf \# spks.}: Total number of speakers. {\bf \# spk combs.}: Total number of speaker combinations. {\bf \# pairs}: Total number of utterance pairs. {\bf \# ratings}: Total number of similarity ratings.}
    \vspace{-2mm}
    \resizebox{0.95\linewidth}{!}{
        \begin{tabular}{ccccc}
        \toprule
        \textbf{Set} & \textbf{\# spks.} & \textbf{\# spk combs.} & \textbf{\# pairs} & \textbf{\# ratings} \\
        \midrule
        Train & 1,142 & 24,764 & 38,802 & 63,845 \\
        Test & 109 & 904 & 2,776 & 5,564 \\
        \midrule
        Total & 1,251 & 25,668 & 41,578 & 69,409 \\
        \bottomrule
        \end{tabular}
    }
    \vspace{-3mm}
    \label{tab:data_split}
\end{table}

\vspace{3pt}\noindent\textbf{Data statistics.} 
The total number of speakers in the collected data is 1,251, which matches the total number of speakers in VoxCeleb1. Train and test sets are structured to include distinct speakers, thereby ensuring the trained models' generalisation capabilities for unseen speakers. This division results in 1,142 speakers for training and 109 speakers for testing. After segregating the speakers, refining the collected scores yields 69,409 ratings for 41,578 pairs. In summary,  dataset statistics and the train/test split are provided in Table~\ref{tab:data_split}, and the distributions of scores are illustrated in Fig.~\ref{fig:sdist}.

\begin{table*}[!t]
    \centering
    \caption{VoxSim test set results. {\rm pt.:} pre-train. {\rm ft.:} fine-tune.}
    \vspace{-2mm}
    \resizebox{0.9\linewidth}{!}{
        \begin{tabular}{lccccc}
        \toprule
        \textbf{Model} & \textbf{LCC} $\uparrow$ & \textbf{SRCC} $\uparrow$ & \textbf{R2} $\uparrow$ & \textbf{MSE} $\downarrow$ & \textbf{ACC} $\uparrow$ \\
        \midrule
        ECAPA-TDNN & & & & & \\
        \quad pt. speaker recogniser & 0.768 & 0.758 & 0.521 & 1.471 & 0.316 \\
        \qquad $\drsh$ ft. on individual scores & 0.827 {\scriptsize $\pm 0.002$} & 0.824 {\scriptsize $\pm 0.004$} & 0.681 {\scriptsize $\pm 0.003$} & 0.981 {\scriptsize $\pm 0.008$} & 0.412 {\scriptsize $\pm 0.003$} \\
        \qquad $\drsh$ ft. on average scores & \textbf{0.829} {\scriptsize $\pm 0.001$} & \textbf{0.828} {\scriptsize $\pm 0.001$} &\textbf{ 0.685} {\scriptsize $\pm 0.001$} & \textbf{0.967} {\scriptsize $\pm 0.002$} & \textbf{0.419} {\scriptsize $\pm 0.006$} \\
        \midrule
        WavLM-ECAPA & & & & & \\
        \quad pt. speaker recogniser & 0.752 & 0.736 & 0.505 & 1.520 & 0.306 \\
        \qquad $\drsh$ ft. on individual scores & 0.833 {\scriptsize $\pm 0.001$}  & 0.835 {\scriptsize $\pm 0.000$} & 0.690 {\scriptsize $\pm 0.002$} & 0.951 {\scriptsize $\pm 0.005$} & 0.402 {\scriptsize $\pm 0.007$} \\
        \qquad $\drsh$ ft. on average scores & \textbf{0.835} {\scriptsize $\pm 0.002$} & \textbf{0.836} {\scriptsize $\pm 0.001$} & \textbf{0.693} {\scriptsize $\pm 0.003$} & \textbf{0.943} {\scriptsize $\pm 0.010$} & \textbf{0.405} {\scriptsize $\pm 0.004$} \\
        \midrule
        SVSNet & & & & & \\
        \quad train on individual scores & \textbf{0.758} {\scriptsize $\pm 0.001$} & \textbf{0.753} {\scriptsize $\pm 0.002$} & \textbf{0.549} {\scriptsize $\pm 0.003$} & \textbf{1.384} {\scriptsize $\pm 0.009$} & \textbf{0.397} {\scriptsize $\pm 0.006$} \\
        \quad train on average scores & 0.747 {\scriptsize $\pm 0.006$} & 0.742 {\scriptsize $\pm 0.006$} & 0.530 {\scriptsize $\pm 0.018$} & 1.443 {\scriptsize $\pm 0.054$} & 0.378 {\scriptsize $\pm 0.005$} \\
        \bottomrule
        \end{tabular}
    }
    \vspace{-1mm}
    \label{tab:voxsimtest}
\end{table*}
\section{Experimental Setup}

\subsection{Model architectures}
We adopt three model architectures for speaker similarity prediction experiments. ECAPA-TDNN~\cite{desplanques2020ecapa} is a state-of-the-art model designed for automatic speaker verification. Given the attempts~\cite{cooper2022generalization, maniati22somos} to apply self-supervised learning (SSL) based models to develop Mean Opinion Scores (MOS) prediction models to enhance generalisation performance across multiple speech datasets, WavLM-ECAPA is adopted. This model uses SSL-based WavLM~\cite{chen2022wavlm} as a feature encoder. Finally, we also experiment with SVSNet~\cite{hu2022svsnet}, the only publicly available model specifically designed for predicting speaker similarity.

\vspace{3pt}\noindent\textbf{ECAPA-TDNN.}
ECAPA-TDNN enhances the traditional Time Delay Neural Network (TDNN) architecture by incorporating Squeeze-and-Excitation (SE) blocks to recalibrate channel-wise feature responses dynamically. Additionally, the model employs a multi-layer feature aggregation mechanism that enhances its ability to capture speaker characteristics across various temporal resolutions. This architecture leverages the power of attention mechanisms and convolutional layers to achieve superior performance in speaker verification. 

\vspace{3pt}\noindent\textbf{WavLM-ECAPA.}
In Chen et al.~\cite{chen2022large}, the authors leverage speech representations extracted from SSL-based pre-trained models for automatic speaker verification. The integration of a pre-trained feature encoder and ECAPA-TDNN downstream network demonstrates significant improvement in verification performance. In our experiments, we use WavLM as the feature extraction model, which has shown strong performance in several speech processing tasks. WavLM is trained with masked speech denoising and prediction in the pre-training, which makes it robust in complex acoustic environments and effective in preserving speaker identity.

\vspace{3pt}\noindent\textbf{SVSNet.} 
SVSNet takes raw waveforms as input to fully utilise speech information for prediction and aligns the representations of the two inputs in two directions through a co-attention module. The model can be trained either in a regression manner using an L2 loss or in a classification manner using a cross-entropy loss, with the regression approach being shown to be superior.

\subsection{Implementation details}
Our implementation is based on the PyTorch~\cite{paszke2019pytorch} framework and is trained on an NVIDIA RTX A6000 with 48GB of memory. During the training of ECAPA-TDNN and WavLM-ECAPA, a 3-second segment from each utterance is randomly sampled to form a batch. Additionally, for ECAPA-TDNN, an 80-dimensional filterbank is extracted as input. 
These models extract a 256-dimensional speaker embedding for each utterance and predict speaker similarity by computing the cosine similarity between the extracted embeddings. The predicted score is compared to a similarity label projected on a 0 to 1 scale, and the model is trained using MSE loss.
To facilitate effective speaker feature extraction at the beginning of training, we use models pre-trained with a speaker identification approach on the VoxCeleb dataset. An Adam~\cite{kingma2014adam} optimizer is employed with an initial learning rate of $10^{-5}$, which decreases by 5\% at each epoch. For SVSNet, a regression-based model is adopted, following the experimental setup of the original paper~\cite{hu2022svsnet}\footnote{https://github.com/n1243645679976/SVSNet}. Every experiment is conducted three times independently to reduce the impact of random initialisation, and the average and standard deviation of these experiments are reported.

\subsection{Evaluation metrics}
The evaluation is based on the model's predicted similarity score compared to the average similarity score of the utterance pair. The model is evaluated with the following metrics: linear correlation coefficient (LCC), Spearman's rank correlation coefficient (SRCC), coefficient of determination (R2), mean squared error (MSE), and accuracy (ACC). For accuracy, a prediction is considered a true positive if the predicted similarity score is within 0.5 from the ground-truth label.

\section{Results}
\subsection{Results on VoxSim test set}
We first compare ECAPA-TDNN, WavLM-ECAPA, and SVSNet on the VoxSim test set, training each model using either individual scores or average scores per utterance pair. ECAPA-TDNN and WavLM-ECAPA are fine-tuned from a pre-trained speaker recognition model, whereas SVSNet is trained from scratch as it is not specifically designed for speaker recognition. The experimental results are summarised in Table~\ref{tab:voxsimtest}. Although the pre-trained speaker recogniser ECAPA-TDNN and WavLM-ECAPA exhibit strong performance on VoxCeleb1, with speaker verification EERs of 0.96\% and 0.43\%, respectively, they achieve LCCs of only 0.768 and 0.733 for speaker similarity prediction, showing much lower correlation than the models fine-tuned with the VoxSim train set. Notably, WavLM-ECAPA, despite its higher speaker verification performance, shows a lower speaker similarity prediction rate compared to ECAPA-TDNN. This suggests that the speaker recogniser may lose features related to speaker similarity while focusing on distinguishing between different speakers and aggregating embeddings of the same speaker. There is no clear superiority between models trained on individual scores and those trained on average scores. WavLM-ECAPA outperforms the others in all metrics except for accuracy, where ECAPA-TDNN achieves the highest accuracy. 

\begin{table}[!t]
    \centering
    \caption{Speaker recognition pre-training ablation.}
    \vspace{-2mm}
    \resizebox{0.95\linewidth}{!}{
        \begin{tabular}{lcc}
        \toprule
        \textbf{Model} & \textbf{LCC} $\uparrow$ & \textbf{MSE} $\downarrow$ \\
        \midrule
        SVSNet & 0.747 {\scriptsize $\pm 0.001$} & 1.443 {\scriptsize $\pm 0.054$} \\
        \midrule
        ECAPA-TDNN & & \\
        \quad w/o speaker pre-train. & 0.761 {\scriptsize $\pm 0.001$} & 1.297 {\scriptsize $\pm 0.005$} \\
        \quad w/ ~~speaker pre-train. & \textbf{0.829} {\scriptsize $\pm 0.001$} & \textbf{0.967} {\scriptsize $\pm 0.002$} \\
        \midrule
        WavLM-ECAPA & & \\
        \quad w/o speaker pre-train.  & 0.806 {\scriptsize $\pm 0.002$} & 1.090 {\scriptsize $\pm 0.009$} \\
        \quad w/ ~~speaker pre-train. & \textbf{0.835} {\scriptsize $\pm 0.002$} & \textbf{0.943} {\scriptsize $\pm 0.010$} \\
        \bottomrule
        \end{tabular}
    }
    \label{tab:speaker_pt_ablation}
\end{table}
\begin{figure}[!t]
   \centering
   \includegraphics[width=0.99\linewidth]{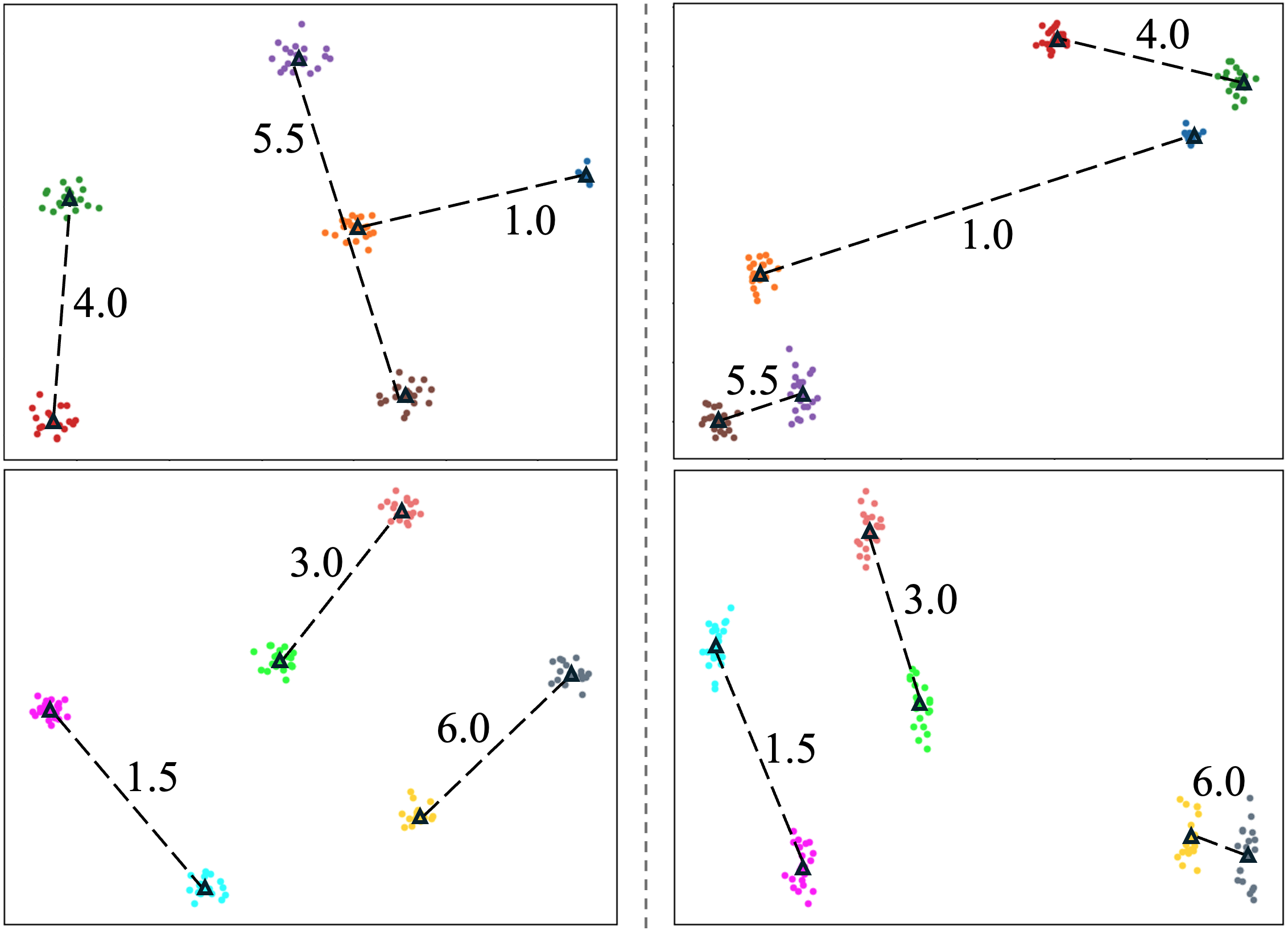}
   \caption{t-SNE plot for embeddings extracted from (left) speaker recognition model and (right) speaker similarity prediction model. Different colours represent distinct speakers, and the number next to the dashed line represents the human-assessed speaker similarity between two utterances.}
   \label{fig:tsne}
\end{figure}

\vspace{4pt}\noindent\textbf{Speaker recognition pre-training ablation.} 
In particular, the fine-tuned ECAPA-TDNN and WavLM-ECAPA exhibit significant predictive performance compared to SVSNet, which can be attributed to the speaker recognition pre-training. To establish the effectiveness of speaker recognition pre-training, we train ECAPA-TDNN and WavLM-ECAPA from scratch. As shown in Table~\ref{tab:speaker_pt_ablation}, both models perform similarly to SVSNet when trained without speaker recognition pre-training. However, this pre-training improves LCC by 8.9\% and 3.6\%, respectively. This demonstrates that speaker recognition pre-training significantly enhances speaker similarity prediction performance.

\vspace{4pt}\noindent\textbf{Qualitative results from t-SNE plot.} 
To demonstrate that the speaker similarity prediction model fine-tuned on VoxSim captures perceptual similarity, we visualise the speaker embeddings extracted from the model using t-SNE~\cite{van2008visualizing} plots. To clearly show the effect of training, plots for embeddings extracted from the pre-trained speaker recognition model are also provided for comparison. As illustrated in Fig.~\ref{fig:tsne}, the speaker recognition model separates different speaker embedding clusters far apart regardless of perceived speaker similarity, whereas the similarity prediction model accurately reflects perceived speaker similarity. Furthermore, each speaker cluster in the speaker similarity prediction model exhibits soft boundaries and is relatively spread out. This indicates that the model has learnt that human-perceived speaker characteristics vary depending on the acoustic environment of the utterance.

\subsection{Results on VCC2018 test set}
To verify the utility of our dataset, we test the generalisability of the models trained on VoxSim using the VCC2018 dataset, which contains natural and synthesised speech from voice conversion systems. Experiments are conducted in both zero-shot and fine-tuning settings, using the same train and test sets as those in the original SVSNet~\cite{hu2022svsnet} paper. Table 5 shows the results on the VCC2018 test set. The fine-tuning results show that all models pre-trained with VoxSim outperform those trained solely on VCC2018. ECAPA-TDNN and WavLM-ECAPA with VoxSim pre-training demonstrate LCC improvements of 5.0\% and 2.5\%, respectively, over models trained without pre-training. Interestingly, models trained solely on VoxSim perform similarly to those trained on VCC2018, with scores of 0.576 vs. 0.562 for ECAPA-TDNN and 0.594 vs. 0.566 for WavLM-ECAPA. This demonstrates the excellent generalisability of models trained with VoxSim.

\begin{table}[!t]
    \centering
    \caption{VCC2018 test set results.}
    \vspace{-2mm}
    \resizebox{0.95\linewidth}{!}{
        \begin{tabular}{lcc}
        \toprule
        \textbf{Model} & \textbf{LCC} $\uparrow$ & \textbf{MSE} $\downarrow$ \\
        \midrule
        ECAPA-TDNN & & \\
        \quad pt. speaker recogniser & 0.512 & 1.090 \\
        \qquad $\drsh$ ft. on VCC2018 & 0.576 {\scriptsize $\pm 0.001$} & 0.862 {\scriptsize $\pm 0.000$} \\
        \quad pt. on VoxSim & 0.562 {\scriptsize $\pm 0.004$} & 0.901 {\scriptsize $\pm 0.010$} \\
        \qquad $\drsh$ ft. on VCC2018 & \textbf{0.605} {\scriptsize $\pm 0.000$} & \textbf{0.806} {\scriptsize $\pm 0.001$} \\
        \midrule
        WavLM-ECAPA & & \\
        \quad pt. speaker recogniser & 0.439 & 1.286 \\
        \qquad $\drsh$ ft. on VCC2018 & 0.594 {\scriptsize $\pm 0.000$} & 0.828 {\scriptsize $\pm 0.002$} \\
        \quad pt. on VoxSim & 0.566 {\scriptsize $\pm 0.003$} & 0.884 {\scriptsize $\pm 0.004$} \\
        \qquad $\drsh$ ft. on VCC2018 & \textbf{0.609} {\scriptsize $\pm 0.000$} & \textbf{0.800} {\scriptsize $\pm 0.000$} \\
        \midrule
        SVSNet & & \\
        \quad SVSNet~\cite{hu2022svsnet} & 0.574 & 0.844 \\
        \quad SVSNet (Ours) & 0.575 {\scriptsize $\pm 0.001$} & 0.849 {\scriptsize $\pm 0.003$} \\
        \quad pt. on VoxSim & 0.509 {\scriptsize $\pm 0.004$} & 3.237 {\scriptsize $\pm 0.266$} \\
        \qquad $\drsh$ ft. on VCC2018 & \textbf{0.586} {\scriptsize $\pm 0.001$} & \textbf{0.839} {\scriptsize $\pm 0.006$} \\
        \bottomrule
        \end{tabular}
    }
    \vspace{-1mm}
    \label{tab:vcc_test}
\end{table}
\section{Conclusion}
We present VoxSim, a speaker similarity evaluation dataset featuring nearly 70k scores for 41k pairs of utterances. This is the first large-scale dataset specifically collected to develop speaker similarity prediction models. Our dataset includes 1,251 speakers, and the utterances span a wide range of acoustic environments and contents. We collect speaker similarity scores from 12 evaluators through listening tests and document a detailed test design for data reliability. We provide baseline results for three speaker similarity prediction model architectures on VoxSim and demonstrate their generalisability through zero-shot and fine-tuning experiments on VCC2018 data.
\section{Acknowledgement}
This work was supported by Samsung Research.

\bibliographystyle{IEEEtran}
\bibliography{main}

\end{document}